\documentclass[conference]{IEEEtran}
\IEEEoverridecommandlockouts
\usepackage{cite}
\usepackage{amsmath,amssymb,amsfonts}
\usepackage{algorithmic}
\usepackage{graphicx}
\usepackage{textcomp}
\usepackage{xcolor}
\usepackage{amsmath,graphicx}
\usepackage{epstopdf}
\usepackage{array}
\usepackage{graphicx}
\usepackage{multirow}
\usepackage{color, soul}
\usepackage{cite}
\usepackage{url}
\usepackage{hyperref}
\usepackage{booktabs}
\usepackage{amsmath}
\usepackage{amssymb}

\def\BibTeX{{\rm B\kern-.05em{\sc i\kern-.025em b}\kern-.08em
    T\kern-.1667em\lower.7ex\hbox{E}\kern-.125emX}}
\begin{document}

\title{MnTTS: An Open-Source Mongolian Text-to-Speech Synthesis Dataset\\ and Accompanied Baseline}

\author{
\IEEEauthorblockN{Yifan Hu$^\dagger$, Pengkai Yin$^\dagger$, Rui Liu$^*$\thanks{$^\dagger$: Equal contribution.}
\thanks{$^*$: Corresponding author is Rui Liu.}
\thanks{This work was supported by the High-level Talents Introduction Project of Inner Mongolia University (No. 10000-22311201/002)  and the Young Scientists Fund of the National Natural Science Foundation of China (NSFC) (No. 62206136).}, Feilong Bao and Guanglai Gao }
\centerline~\IEEEauthorblockA{ \textit{College of Computer Science, Inner Mongolia University, Hohhot, China
}\\ 
Email: hyfwalker@163.com, yinpengkai1933@163.com, liurui\_imu@163.comm, \{csfeilong, csggl\}@imu.edu.cn}
}


\maketitle

\begin{abstract}
This paper introduces a high-quality open-source text-to-speech (TTS) synthesis dataset for Mongolian, a low-resource language spoken by over 10 million people worldwide. The dataset, named MnTTS, consists of about 8 hours of transcribed audio recordings spoken by a 22-year-old professional female Mongolian announcer.
It is the first publicly available dataset developed to promote Mongolian TTS applications in both academia and industry. In this paper, we share our experience by describing the dataset development procedures and faced challenges. To demonstrate the reliability of our dataset, we built a powerful non-autoregressive baseline system based on FastSpeech2 model and HiFi-GAN vocoder, and evaluated it using the subjective mean opinion score (MOS) and real time factor (RTF) metrics. Evaluation results show that the powerful baseline system trained on our dataset achieves MOS above 4 and RTF about $3.30\times10^{-1}$, which makes it applicable for practical use. The dataset, training recipe, and pretrained TTS models are freely available \footnote{\label{github}\url{https://github.com/walker-hyf/MnTTS}}.
\end{abstract}


\section{Introduction}
Text-to-speech (TTS) aims to synthesize high-quality speech for any given text \cite{taylor2009text}. TTS is an important research direction in artificial intelligence (AI) and received wide attention from academia and industry \cite{tan2021survey}. It has a wide range of applications, such as navigation, announcement, smart assistants, and other speech-enabled devices. 
With the development of deep learning technology, high-quality training data has become a necessary condition for training a reliable neural network model. Therefore, to build a robust TTS system, a high-quality speech dataset is required. 
For the mainstream languages such as Chinese and English, there are a lot of large-scale high-quality speech data, such as LJSpeech \cite{ito2017lj}, libriTTS\cite{zen2019libritts}, AiShell\cite{shi2020aishell}, etc. However, for some low-resource languages such as Mongolian, such data is scarce.
In order to address this, we developed a open-source speech dataset for the Mongolian language.
We named our dataset MnTTS, and it is primarily geared to build high-quality Mongolian TTS systems.

Mongolian language belongs to the Mongolian branch of the Mongolian language family of the Altai language family, which is the most famous and widely spoken language of the Mongolian language family. Mongolian is mainly used in Mongolian-inhabited areas of China, Mongolia and the Siberian Federal District of the Russian Federation.
At the same time, Mongolian is also the main national language in Inner Mongolia Autonomous Region of China. 
In the world, the number of speakers is about 6 million~\cite{puthuval2017language}. Furthermore, there is a growing awareness of the importance of increasing the number of Mongolian speakers reflected in many language rescue initiatives \footnote{\url{https://news.mn/en/791396/}} launched by the government.
Therefore, the study of speech synthesis technology for Mongolian is of great significance for education, transportation, communication and other fields in ethnic minority areas.
Note that the Mongolian language used in Mongolia country is mainly spelled in Cyrillic scripts~\cite{cohen2005english} because of the influence of the former Soviet Union in the 1950s and 1960s, while used in China is mainly spelled in traditional scripts~\cite{rohsenow2004fifty}. This paper focus on the traditional scripts.



Currently, there is no Mongolian speech dataset of sufficient quality for building TTS systems, especially recently proposed end-to-end (E2E) neural architectures \cite{wang2017tacotron,sotelo2017char2wav,arik2017deep,skerry2018towards,shen2018natural,ren2019fastspeech,ren2020fastspeech,2018Neural}, such as Tacotron\cite{wang2017tacotron}, Tacotron2 \cite{shen2018natural} models. 
Armed with WaveNet-like vocoders, the effect of synthetic speech has reached the level of human pronunciation. To further speed up the inference process, non-autoregressive TTS models and vocoders, like FastSpeech \cite{ren2019fastspeech}, FastSpeech2 \cite{ren2020fastspeech}, MelGAN \cite{kumar2019melgan}, VocGAN \cite{yang2020vocgan}, HiFi-GAN \cite{kong2020hifi} etc., are proposed and achieved excellent performance.
This work aims to fill the gap for Mongolian by introducing the MnTTS dataset. To the best of our knowledge, it is the first open-source dataset developed for building Mongolian TTS systems. Our dataset contains around 8 hours of high-quality speech data read by a 22-year-old professional female Mongolian announcer. The dataset was carefully annotated by native transcribers and includes politics, business, sports, entertainment, and other fields. The manuscript covers all Mongolian alphabets and rich word combinations. The MnTTS is freely available for both academic and commercial use under the Creative Commons Attribution 4.0 International License\footnote{\label{licenses}\url{https://creativecommons.org/licenses/by/4.0/}}.

By introducing the MnTTS database, we plan to promote the development of Mongolian intelligent information processing technology, which will play an important role in promoting the development of artificial intelligence technology for ethnic minorities in China. We believe that the MnTTS database will be a valuable resource for the TTS research community, and our experience will benefit other researchers planning to develop speech datasets for low-resource languages. We note that the primary application domain of MnTTS is speech synthesis. However, we believe that our data will also be useful for speech recognition, speech enhancement and other related fields.

To demonstrate the reliability of MnTTS, we combine the Fastspeech2 \cite{ren2020fastspeech} model and HiFi-GAN \cite{kong2020hifi} vocoder to build our baseline system since the FastSpeech2 is the state-of-the-art non-autoregressive acoustic model and HiFi-GAN is the state-of-the-art non-autoregressive vocoder. 
We evaluated the system using the subjective mean opinion score (MOS) and real time factor (RTF) measures. 
The experiment results showed that the TTS model built using our dataset achieve 4.46 in MOS and $3.30\times10^{-1}$ in RTF for the female speaker, respectively, which assures the usability for practical applications. 
In addition, we performed an analysis on the robustness issue. Note that we found some unstable phenomena such as word skipping and word missing in the synthetic examples, and we conducted an in-depth analysis of the reasons.
At last, the MnTTS dataset, training recipe, and pretrained models are publicly available~\ref{github}.

The rest of the paper is organized as follows. Section II briefly reviews related works. Section III describes the MnTTS construction procedures and reports the statistics information. Section IV explains the TTS experimental setup and discusses obtained results. Section V concludes the paper and highlights future research directions. 

\section{Related works}

\subsection{TTS dataset construction}
The recent proliferation of human-machine interaction applications, such as smart speaker, smart home, smart car assistant, has attracted a great deal of attention to TTS research from both academia and industry~\cite{wang2017tacotron,sotelo2017char2wav,arik2017deep}. 
Consequently, many large-scale datasets, such as LJSpeech \cite{ren2019fastspeech}, libriTTS\cite{zen2019libritts}, AiShell\cite{shi2020aishell}, etc., have been collected and released freely \cite{ito2017lj,zen2019libritts,shi2020aishell,veaux2016superseded}.
However, these datasets are mostly focus on resource-rich languages, such as English, Mandarin, and so on.

To build a TTS dataset for low-resource language, some researchers proposed unsupervised \cite{ren2019almost}, semi-supervised \cite{chung2019semi}, and cross-lingual transfer learning \cite{tu2019end} based methods. 
Specifically, these methods first crawl some voice files automatically from the Internet, then adopt speech recognition models or extract some language knowledge from another similar language to label the raw audio files. 
Although the above methods provide some available speech corpora to support TTS model training and achieve acceptable speech synthesis performance, the overall quality of the synthesized speech is usually insufficient for practical applications due to the environmental noise and background noise etc.. 

We note that the simplest and most straightforward idea is to record and manually annotate audio recordings.
Although such a process is notoriously laborious and time consuming. However, in order to promote the development of low resource language TTS, this method is a reliable method to obtain high quality speech synthesis results.

In this paper, we will adopt the second approach to prepare our MnTTS dataset, and we believe that it is worthwhile to open source such high-quality Mongolian speech dataset.

\subsection{Mongolian TTS}

The study of Mongolian speech synthesis has a long history. In recent years, with the development of deep learning technology, the research on Mongolian speech synthesis has ushered in a new climax.

In \cite{liu2020exploiting}, deep learning techniques were first introduced to Mongolian speech synthesis, and DNN-based acoustic models trained on 5-hours training data were used instead of HMM acoustic models to improve the overall performance of Mongolian TTS.
In \cite{8706263}, a Tacotron-based Mongolian speech synthesis system trained on about 17 hours of training data was implemented. 
Among them, the overall performance of the tacotron-based Mongolian TTS system achieved significant improvement compared with the traditional methods.
Similarly, Huang et al.\cite{9675192} realized Mongolian emotional speech synthesis based on transfer learning and emotional embedding, but their dataset was not shared also. 

The above mentioned works provide a solid foundation for the research of Mongolian TTS technology. However, the datasets involved are not publicly available. 
Therefore, it is necessary to build a high-quality open-source Mongolian TTS dataset, which will be the focus of this work.

 \section{MnTTS dataset}

The MnTTS project was conducted with the approval of the xx lab. The female speaker participated voluntarily and was informed of the data collection and use protocols through a consent form.

\subsection{Text collection and narration}
{The first step in building this dataset is to conduct text collection. For the scope of text collection, we initially selected mainstream platforms such as news websites, social media platforms, and books. In terms of topics, we will try to ensure that the selected texts have a wide coverage (e.g., politics, business, sports, entertainment, culture, etc.). We also use manual filtering to exclude inappropriate content, such as more sensitive political issues or issues related to privacy and violent pornography. Under this rule, we collected a total of 7,000 utterances as our final text scripts.}


\begin{table}[]
\centering
\caption{\label{tab:statics}The MnTTS dataset specifications}
\begin{tabular}{p{2cm}<{\centering}p{2cm}<{\centering}p{2cm}<{\centering}}
\toprule[1pt]
Category & \multicolumn{2}{c}{Statistics} \\ \hline
\multirow{4}{*}{Character} & Total & 410044 \\
 & Mean & 66 \\
 & Min & 2 \\
 & Max & 189 \\ \hline
\multirow{4}{*}{Phoneme} & Total & 310565 \\
 & Mean & 50 \\
 & Min & 1 \\
 & Max & 146 \\ \hline
\multirow{4}{*}{Word} & Total & 63866 \\
 & Mean & 10 \\
 & Min & 1 \\
 & Max & 28 \\ \bottomrule[1pt]
\multicolumn{1}{l}{} & \multicolumn{1}{l}{} & \multicolumn{1}{l}{}
\end{tabular}
\end{table}

\subsection{Text preprocessing}

The traditional Mongolian language has a unique agglutinative language characteristic, which makes the expression of Mongolian letters in words variable, and its manifestation may vary in different contexts.
Therefore, there is a serious phenomenon of homophony that lead to many incorrectly encoded letters in text scripts. 
To avoid such a problem as much as possible, we convert the Mongolian text to a Latin sequence representation. The text is processed in three steps: encoding correction, Latin conversion, and text regularization.

\begin{itemize}
    \item \textbf{Encoding correction}: Firstly, we manually convert the incorrect encoding to the correct form \cite{liu2018phonologically} to correct the Mongolian character encoding.

    \item \textbf{Latin conversion}: After that,  the corrected Mongolian characters are converted to the Latin representation according to the Alphabet-Latin mapping table \cite{liu2019building}.

    \item \textbf{Text regularization}: Finally, more than 140 \cite{liu2019building} kinds of regular expressions were designed to filter the special characters that appear in Mongolian with high frequency, such as date and Arabic numerals, to convert the irregular Mongolian text into the standardized Mongolian Latin character representation sequence.
\end{itemize}

\subsection{Audio recording and audio-text alignment}

In order to ensure the quality of the audio recording, we invited a 22-year-old professional Mongolian announcer who is a native Mongolian-speaking girl.
The whole recording process was done in a standard recording studio of xx university, and we used \textit{Adobe Audition} software for voice recording. 
The announcer follows our text script and reads it sentence by sentence.
In addition, there is another volunteer to supervise the recording process and re-record if there are any problems, such as murmurs and unreasonable pauses in the recording.

\begin{figure}[]
\centering
\centerline{ \quad \quad \includegraphics[width=\linewidth]{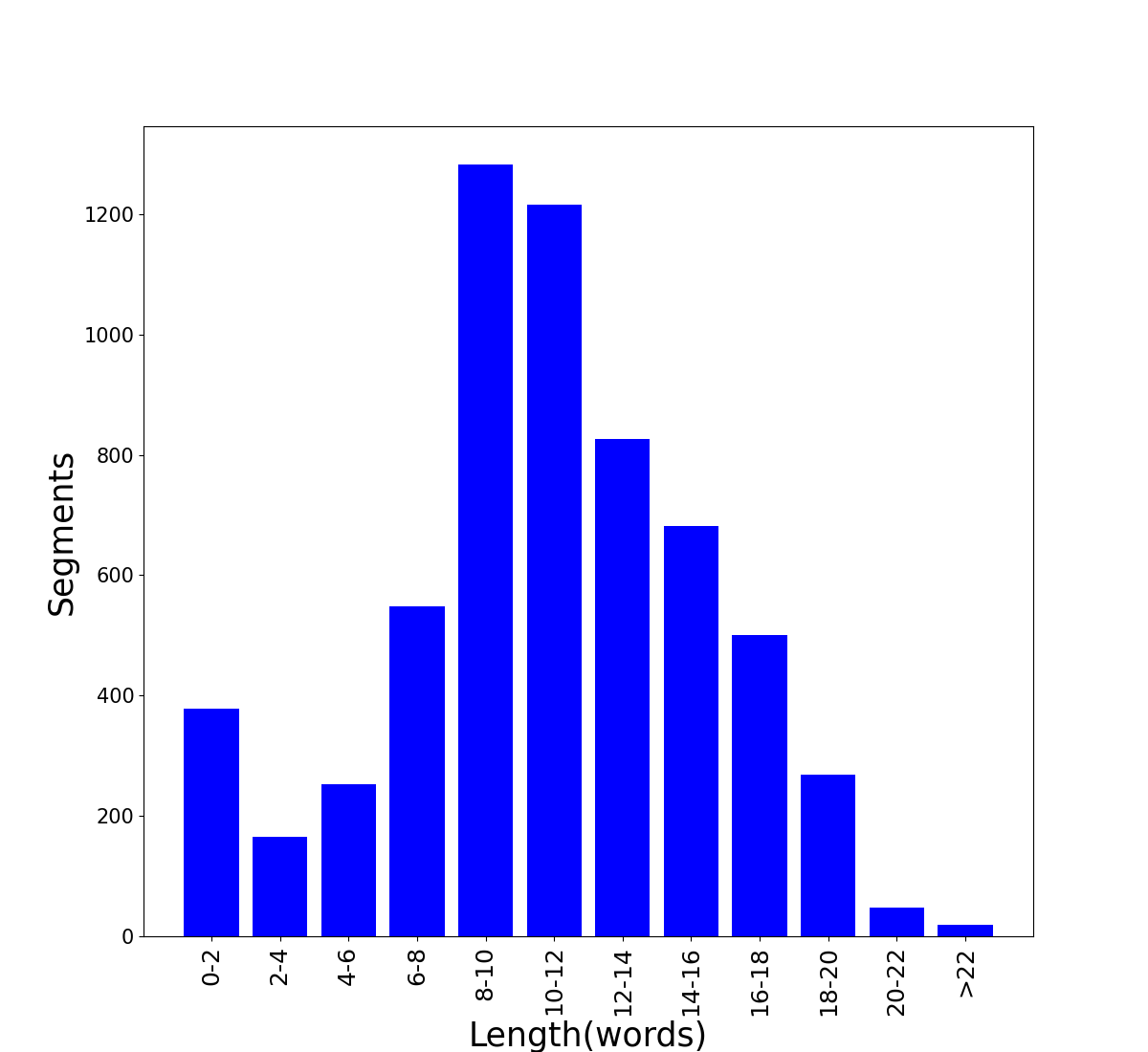}}
\caption{Word length statistics of the MnTTS dataset.}
\label{fig:word}
\vspace{-5mm}
\end{figure}

\begin{figure}[]
 \centering
\centerline{ \quad \quad \includegraphics[width=0.95\linewidth]{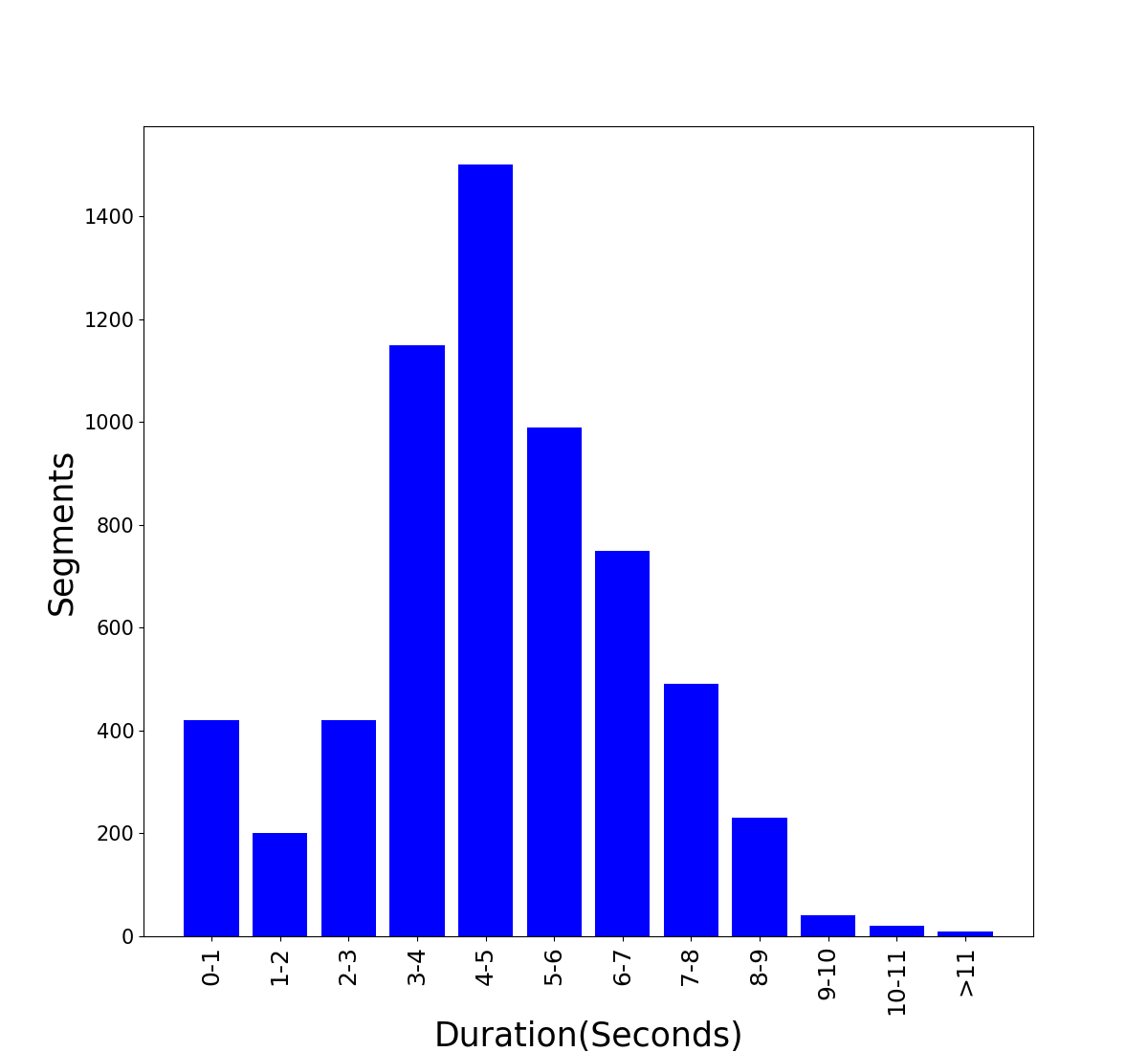}}
\vspace{-2mm}
\caption{Sentence duration statistics of the MnTTS dataset.}
\label{fig:duration}
 \end{figure}

\begin{figure*}[]
\centering
\centerline{ \quad \quad \includegraphics[width=\linewidth]{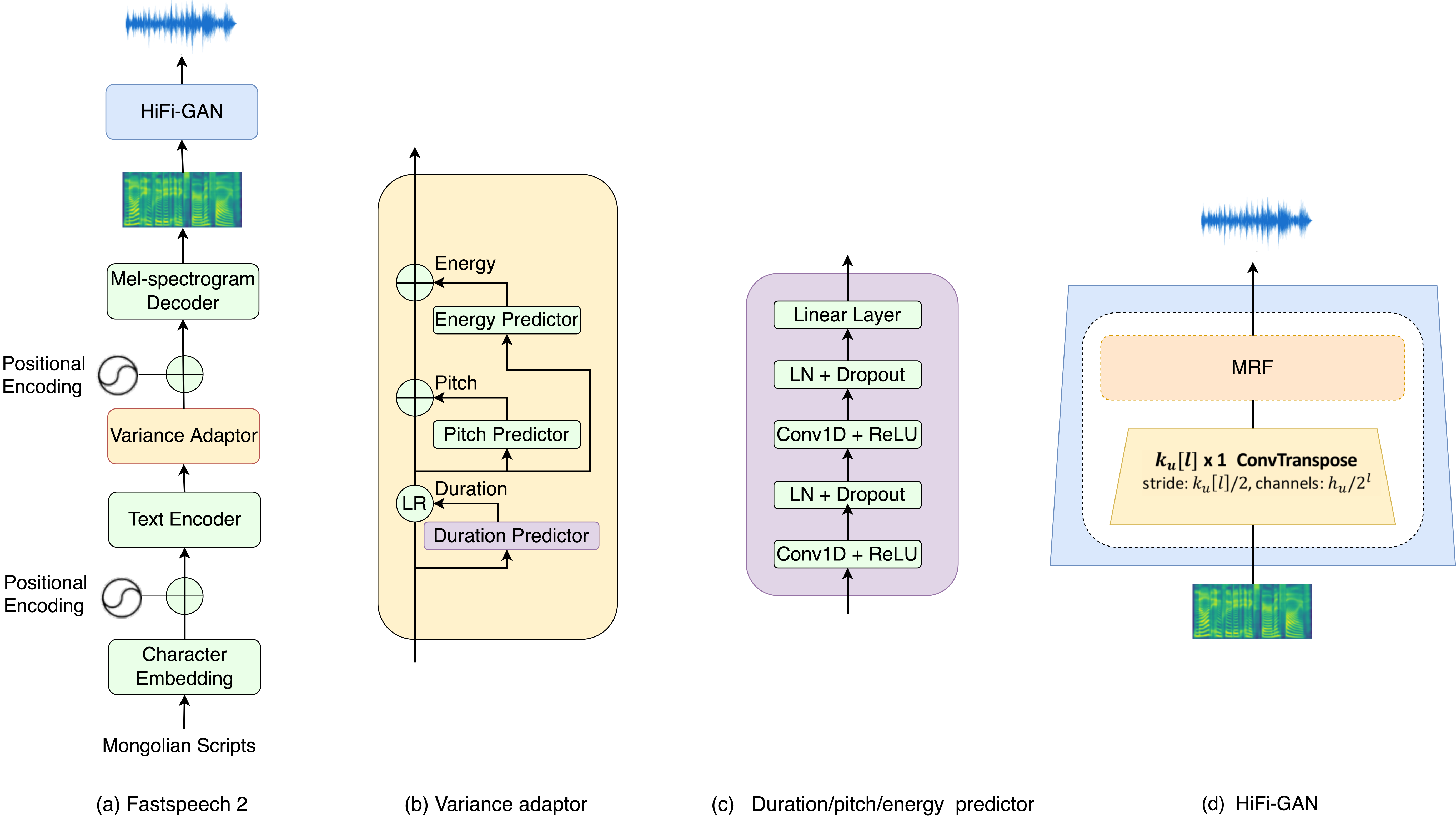}}
\caption{The overall architecture of our baseline system. \textcolor{red}{}}
\label{fig:model}
\end{figure*}

After recording, we also perform further checks on the speech data. Specifically, we invited three Mongolian transcribers to carefully align each sentence with the content of the actual audio. Our textual content consists of Mongolian Latin sequences, referring to each Latin word in the sequence as a word and each letter in the Latin word as a character. the transcripts also contain punctuation marks, such as period (`. ') , comma (`,'), hyphen (`-'), question mark (`?') , exclamation mark (`!') , and so on. We removed excess noise and incorrectly pronounced parts of the audio file, with about 0.3 seconds of silence at the beginning and end of each audio segment. 
In the end, we kept about 6,000 utterances and the corresponding text data.


Finally, we collated about 8 hours of speech data, which was stored in the following format: sampling rate of 44.1 k, sampling precision of 16 bit.

\subsection{Dataset specifications}
The statistical results of the MnTTS data are shown in Table \ref{tab:statics}. The total number of Mongolian characters in the whole data is 410,044, the average (Mean) number of characters per sentence is 66, the shortest (Min) sentence has 2 characters, and the longest (Max) sentence has 189 characters. For phoneme units, this dataset contains 310,565 phoneme units, and we focused on the mean value of 50, the maximum value of 146, and the minimum value of 1 in the statistical process. In addition, we counted the number of words in the data, and the results are shown in the figure, there are 63,866 words in this dataset, the average number of words in a sentence is 10, the shortest sentence has only one word, and the longest has 28 words. And we constructed a histogram Figure \ref{fig:word} using the lengths of the counted sentences, in which we can visually see that most of the sentences are concentrated in the length range of 8-10, and the lengths of the sentences follow a normal distribution.

In terms of sentence duration, as shown in Figure \ref{fig:duration}, we counted the duration of all sentences, and we can see that most of the sentences are concentrated in the duration range of 4-5 seconds, And since this dataset contains a large number of Mongolian names of Mongolian names, which in turn leads to a large portion of the data being concentrated in the 0-1 second interval.

\section{TTS experiments}
To validate the proposed MnTTS dataset, we built the Mongolia TTS system based on FastSpeech2 \cite{ren2020fastspeech} model and HiFi-GAN vocoder and evaluated it using the subjective MOS and RTF measures in terms of naturalness and inference efficiency. Please visit our gitHub repository\ref{github} to enjoy the speech samples.

\subsection{Experimental setup}

The overall architecture of our baseline is shown in the Figure \ref{fig:model}. the FastSpeech2 model aims to convert input Mongolian scripts to the Mel-spectrogram feature, then the HiFi-GAN \cite{kong2020hifi} vocoder seeks to reconstruct the waveform from the Mel-spectrogram feature.
We used TensorFlowTTS toolkit \footnote{\label{tensorflowtts}\url{https://github.com/TensorSpeech/TensorFlowTTS}} to build the baseline system.

FastSpeech2 is the current state-of-the-art non-autoregressive speech synthesis model. Extract duration, pitch, and energy from speech waveforms, use them directly as conditional inputs in training, and use predicted values for inference. This model is not only faster to train, but also alleviates the one-to-many mapping problem in TTS (i.e., multiple phonetic variants corresponding to the same text). The encoder hidden size is set to 384 and the number of hidden layers is 4. The decoder hidden size is 384 and the number of hidden layers is 4. The number of convolutional layers of the predictor in the Variance adaptor is set to 2, and the predictor dropout rate is 0.5. The initial learning rate is 0.001, and the hidden dropout rate is set to 0.2.

HiFi-GAN is a vocoder that is commonly used in both academia and industry in recent years. It can convert the spectrum generated by the acoustic model into high-quality audio. This vocoder uses a generative adversarial network as the basic generative model. The generator of HiFi-GAN mainly has two parts, one is the upsampling structure, which is composed of one-dimensional transposed convolution; the other is the multi-receptive field fusion module, which is mainly responsible for optimizing the sampling points obtained by the upsampling. poor network composition. There are two discriminators of HiFi-GAN, namely multi-scale and multi-period discriminators, which identify speech from two different angles. For generator, kernel size is 7, upsample scales is [8,8,2,2]. For discriminator List of period scales is [2,3,5,7,11], In Conv filters of each period discriminator the filters is 8. For the parameters of melgan discriminator, Pooling type for the input downsampling is AveragePooling1D, List of kernel size is [5,3] , Nonlinear activation function is LeakyReLU.

Before baseline training, a Tacotron2 model trained with 100K steps is used to extract duration from attention alignments for duration predictor of FastSpeech2. 
After that we train Fastspeech2 with 200k steps.
For the HiFi-GAN vocoder, we first train the generator for 100k steps, and then jointly train the generator and discriminator for 200k steps.

All models were trained using the Tesla V100 GPUs. More details on model specifications and training procedures are provided in our GitHub repository\ref{github}.

\begin{table}[]
\centering
\caption{\label{tab:mos}Mean opinion score (MOS) results with 95\% confidence intervals.}
\label{tab:my-table}
\begin{tabular}{ccccc}
\toprule[1pt]
System       & MOS Score &  &  &  \\ \hline
Ground Truth & 4.72 ± 0.03        &  &  &  \\
FastSpeech2 + Griffin-Lim & 4.23 ± 0.03       &  &  &  \\ \hline
\textbf{FastSpeech2 + HiFi-GAN } & \textbf{4.46 ± 
 0.06}        &  &  &  \\ \bottomrule[1pt]

\end{tabular}
\end{table}

\begin{figure}[]
\centering
\centerline{ \quad \quad \includegraphics[width=0.8\linewidth]{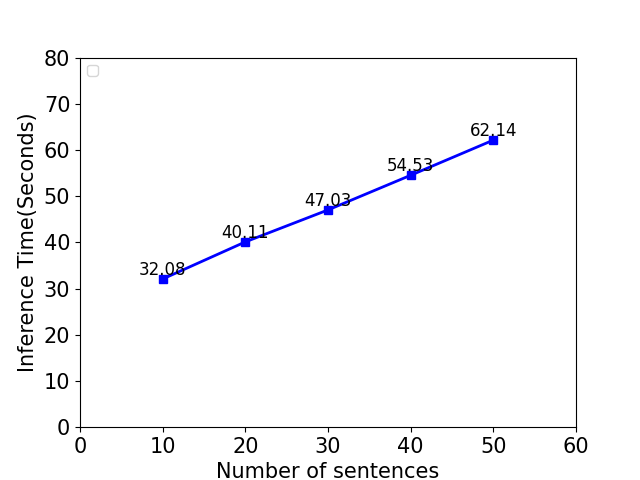}}
\caption{Inference time (seconds) vs. utterance number for our model.}
\label{fig:SPEED}
\end{figure}

\subsection{Naturalness evaluation}

To evaluate naturalness, we conducted the mean opinion score (MOS) test \cite{streijl2016mean}.
As an evaluation set, we randomly select 50 sentences of different lengths. These sentences are not used to train the model. The model-generated audio was randomly shuffled with the ground truth speech and sent along to listeners.
In the test, 5 subjects are asked to use headphones and sit in a quiet environment to rate the naturalness of 250 generated speeches. All subjects were young Mongolian students, with Mongolian as their native language. The recordings were rated using a 5-point Likert scale: 5 for excellent, 4 for good, 3 for fair, 2 for poor, and 1 for bad.

For a full comparison of naturalness, we compared our baseline synthetic speech with the \textbf{Ground Truth} speech. In addition, to verify the performance of HiFi-GAN, we added a \textbf{FastSpeech2+Griffin-Lim} baseline model for further comparison.
Note that the FastSpeech2+Griffin-Lim system adopts the Griffin-Lim \cite{perraudin2013fast} algorithm instead of the HiF-iGAN vocoder to reconstruct the waveform.
The subjective evaluation results are given in Table \ref{tab:mos}. According to the results, the best performance was achieved by the ground truth, as expected.
The worst performer was FastSpeech2 + Griffin-Lim. Importantly, the Fastspeech2 + HiFi-GAN model achieved a MOS measure of above 4.4 and is not too far from the ground truth. This result demonstrates the utility of our MnTTS dataset for TTS applications.

\begin{table}[]
\centering
\caption{Error types found in the 50-sentence test set(total number of words is 500).}
\begin{tabular}{p{5cm}<{\centering}p{3cm}<{\centering}p{3cm}<{\centering}}
\toprule[1pt]
\multicolumn{1}{c}{\multirow{1}{*}{Error types}} & \multicolumn{2}{c}{\multirow{1}{*}{Number}}  \\ \hline
Repeated words & \multicolumn{2}{c}{0} \\
Skipped words & \multicolumn{2}{c}{0} \\
Mispronounced words & \multicolumn{2}{c}{2} \\
Incomplete words & \multicolumn{2}{c}{2} \\
Long pauses & \multicolumn{2}{c}{0} \\
Nonverbal sounds & \multicolumn{2}{c}{1} \\ \hline
\multicolumn{1}{c}{Total} & \multicolumn{2}{c}{5}\\ \bottomrule[1pt]
\label{tab:error}
\end{tabular}
\end{table}

\subsection{Efficiency evaluation}

The same 50-sentence test set from the previous section was used for the efficiency evaluation. The real-time-factor (RTF) metric was calculated by taking the total duration of the 50-sentence test set as the reference \cite{bulut2020low}. Specifically, the time we took to synthesize this test set was 62.14 seconds, and the total duration of the synthesized speech for the test set is 188.48 seconds. Therefore, we obtain the RTF as $3.30\times10^{-1}$ by a division operation.

In addition, we also synthesize different numbers (ranging from 10–50) of sentences and then count the time required. The results are shown in Fig. \ref{fig:SPEED}, from which we find that the inference latency barely increases with the number of sentences for our model. This indicates that our model has good synthesis efficiency and still has fast synthesis speed in batch synthesis, which meets the practical requirements.

\subsection{Robustness analysis}
Although the sound quality of our synthesized voice achieved satisfactory results. However, we also found some unstable synthesis phenomena like skipping words, missing words, etc. during the evaluation.

We identify five types of errors in the synthesized speech of the test set, including repeated words, skipped words, mispronounced words, incomplete words, long pauses, and nonverbal sounds. We invited 5 volunteers to label the error cases for all synthesized speech. Note that the total number of words in the test set is 500. We report the statistical results of the error cases in Table \ref{tab:error}.

From the results, we can see that there are 2 words with mispronunciation, 2 words with incomplete pronunciation, and 1 word with nonverbal noise. 
After analysis, we concluded two reasons. First, the pretrained Tacotron model extracts the duration used to provide a supervised signal for the duration predictor of the FastSpeech2 model, but the tacotron model trained on 8-hours training data may not be robust enough, leading to errors in word duration. Second, HiFi-GAN is also trained based on 8-hours training data, and the vocoder may also be insufficiently trained, leading to noise in some words. In future work, we will continue to improve by increasing the amount of data to address these two issues.

\section{Conclusion}
 
This paper introduced the first open-source Mongolian speech dataset for TTS applications. The MnTTS dataset contains over 8 hours of speech data consisting of around 6,000 recordings. We released our dataset under the Creative Commons Attribution 4.0 International License, which permits both academic and commercial use. We shared our experience by describing the dataset construction and TTS evaluation procedures, which might benefit other researchers planning to collect speech data for other low-resource languages.
To demonstrate the use of our dataset, we built baseline model based on FastSpeech2 and HiFi-GAN vocoder. The MOS and RTF evaluation results suggest that the baseline model trained on MnTTS are suitable for practical use.

In future work, we plan to further extend our dataset by introducing new speakers and emotions. We also plan to explore the optimal hyper-parameter settings for the Mongolian TTS model, compare different TTS architectures, and conduct additional analysis. 


\bibliographystyle{IEEEtran}
\bibliography{refs}

\end{document}